\documentclass[aps,preprint,floatfix,amsmath,superscriptaddress,endfloats,byrevtex]{revtex4}
\usepackage{graphicx}
\usepackage{feynmf}

\begin{document}

\title{Microscopic Computational Model of a Superfluid}

\author{Mikhail Ovchinnikov}
\author{Alexey Novikov}
\affiliation{Department of Chemistry, University of Rochester, RC
Box 270216,
   Rochester, NY 14627-0216, USA}

\date{\today}

\begin{abstract}

A finite one-dimensional microscopic model of a superfulid is presented.
The model consists of interacting Bose particles with an additional impurity
particle confined to a ring.  Both semiclassical and exact quantum calculations
reveal dissipationless motion of impurity with increased effective mass due to its
interaction with the excitations of Bose fluid.  It is shown that both the excitation
spectrum of Bose fluid and the excitation spectrum of impurity can be analyzed using
the structure of the ground state of the system.

\end{abstract}
\maketitle

\section{Introduction}
\label{intro}

Despite its maturity, the theory of Bose-Einstein condensates (BEC) and
superfluidity remains an active field of research.  New theoretical
developments  are motivated by a variety of modern experiments.
Bose Einstein Condensates (BEC) of ultra-cold atoms
in magnetic traps \cite{cornell2002,chikkatur2000} is currently an active
experimental field.   Another set of
recent experiments studies microscopic superfluidity of liquid
helium by spectroscopic measurements on molecules imbedded in
superfluid Helium droplets
\cite{vilesov1998,vilesov2000,dumesh2006}.  It is generally observed
that a microscopic impurity interacting with Bose liquid/gas behaves
as a free particle with an effective mass that is greater than its
original mass. Such dissipationless quantum motion is observed both for
translational motion of particles in a superfluid environment as well as
for rotations of molecules in superfluid helium droplets.
The development of quantum microscopic theory of Bose fluid (BF) / impurity
system has been the theoretical challenge. What is the collective
wavefunction of this system and what is the nature of the effective
mass?   What happens with finite number of Bose particles and how is
the macroscopic limit of superfluidity achieved?  A number of
theoretical works have been devoted to answering those questions.

The system that has been accessible to analytical theory is a
dilute, weakly interacting BEC (see, for example,
\cite{andersen2004} and references therein). In the seminal work of
Bogoliubov \cite{bogol1947} the excitation spectrum and the nature
of excitations of dilute BEC was uncovered. This work set the stage
for most of the further theoretical studies of this system. A number
of authors used perturbation theory to consider the interaction of
an impurity particle with BEC
\cite{girardeau1961,astrak2004,montina2003,miller62,novikov2009spectr}.
It was concluded that the Landau criterion holds for the quantum
motion of a particle in the BEC, {\it i.e.}~the motion of a particle is
free up to a critical momentum.  At low momenta the particle has the
spectrum that resembles that of a free particle, $E(k) \sim k^2$,
thus leading to an effective mass approximation. The effective mass
was calculated by several authors within the Golden Rule limit of
the particle/BEC interaction.  In our recent paper we continued this
work by developing a formal perturbation expansion of this system
based on Coherent State Path Integral formulation of the
particle/BEC dynamics \cite{novikov2009spectr}.  We presented the
diagrammatic representation of perturbation expansion that
allows to evaluate particle properties in dilute BEC up to
an arbitrary order of perturbation theory.  Besides the
perturbation theory another theoretical approach that gained
popularity is the solution of the Gross-Pitaevskii equations for the
BEC \cite{pitaevskii1998}.  Gross-Pitaevskii can be viewed as the
mean field equations of motion for the Bose field. It has been shown that
many of the results obtained from the Gross-Pitaevskii equations,
including dissipation and the effective mass of impurity particles, are the same
as the ones obtained from perturbation theory \cite{astrak2004}.

Despite its important physical insight, the
perturbation theory often cannot be applied to calculate observables
for realistic systems such as liquid helium.  An approach taken by several
theoretical groups has been the computational
imaginary time path integral techniques \cite{ceperley1999}.
The most recent work
has been aimed at the studies of molecular rotations in
superfluid helium  \cite{ceperley2003,whaley2004p,whaley2001k,whaley2004z,
whaley2004zk,whaley2003p,roy2006}.  Whaley and coworkers
\cite{whaley2004p,whaley2001k,whaley2004z,whaley2004zk,whaley2003p}
has calculated the properties of helium droplets with a variety of
molecules.  Roy et al. \cite{roy2006} has considered the limit of
very small helium clusters.  In a number of cases the effective
moments of inertia of molecules were computed and were shown to be in
agreement with experimental values. However, being a
powerful computation tool, this work is necessarily limited to the
calculation of statistical rather than dynamical properties of the
system and does not give a direct answer to the intriguing question of how
does the motion of the molecule and motion of a superfluid uncouple.

A microscopic theory that has been successful in predicting the
excitation spectrum of the bulk superfluid helium originated from the work of
Bijl and Feynman \cite{feenberg,feynman1954}.  In this work the trial
wavefunction of
elementary excitation is built using the unknown wavefunction of the
ground state by the use of the Fourier component of the density
operator, $\psi_k = \hat{\rho}_k \psi_0$.  It is shown that the variational energy of
such state is expressed through the structural properties of the
ground state and is given by $E(k) = k^2/[2mS(k)]$, where $S(k)$
is the superfluid structure factor.  The Bijl-Feynman spectrum has
qualitatively correct shape; however, quantitatively it overestimates
the excitation energy in the most relevant roton minimum spectral region.
Subsequently the theory has been extended to include the multiple Feynman
excitations, an approach generally known as a correlated basis functions
(CBF) theory \cite{feenberg}.  Several realizations of this theory lead to the
excitation spectrum that agrees with the experimental one. Two type of extensions
of the CBF have been applied to
calculate the excitation spectrum of the $^3$He impurity in the superfluid $^4$He.   One
approach is based on the variational method \cite{saarela1993}, while the other is based on
the perturbation series \cite{fabrocini1986,fabrocini1999}.  Both calculations lead to results that
are in quantitative agreement with experiment.

In this work we present a computational study of quantum dynamics of
a very simple model system.  The developed model retains many
important features of the Bose superfluid, it is not limited to the
low density and weak interactions, yet it is small enough that its
quantum dynamics can be computed and analyzed exactly.  The
calculations presented below clearly show the effects of microscopic
superfluidity in a finite, one-dimensional system.  We analyze the
results using the analytical theory developed for BEC and superfluid
helium.  The work provides an extensive illustration of these
theories and allows to clearly understand the limits of their
applicability.  In the next Section the Hamiltonian of the model is
presented.  In Section III the semi-classical equations of motion
for the model are solved.  It is shown that the model exhibits
microscopic superfluidity and the nature of effective mass is
revealed.  In Section IV the exact numerical solution for a pure BF
is presented.  We find the excitation spectrum of this system and
compare it to the Bogoliubov's spectrum.  In Section V the
free-particle like excitation spectrum of impurity is obtained.  Finally,
in Section VI we show that the approach based on the Bijl-Feynman
theory captures the essential physics of the impurity excitation
spectrum and provides an accurate method for calculating the
effective mass of impurity based on the structural properties of the
ground state.

\section{Model}

The main object in our model is one-dimensional BF.  In order to
make it accessible for numerical calculations the system is made
finite by confining the BF to a ring of radius R (enforcing periodic
boundary conditions).  The system schematically represented in Fig.~1.  The Bose
particles represented by empty circles interact by pair potential
$U(x_i-x_j)$, where $x_i$ and $x_j$ are the coordinates of
the particles on a ring, the difference $x_i-x_j$ is the shortest
distance between particles.  An impurity which
interacts with BF via the pair potential $V(x_i - x)$,
where $x$ is the coordinate of an impurity and $x_i$ is the
coordinate of $i$th Bose particle, can be added to this system.
For the sake of computational
simplicity the interaction potentials are taken to be Gaussians of unit widths,
$U(x) = \alpha e^{-x^2}$, and $V(x) = \beta e^{-x^2}$.

\begin{figure}
\label{model_fig}
\includegraphics[width=17.0cm]{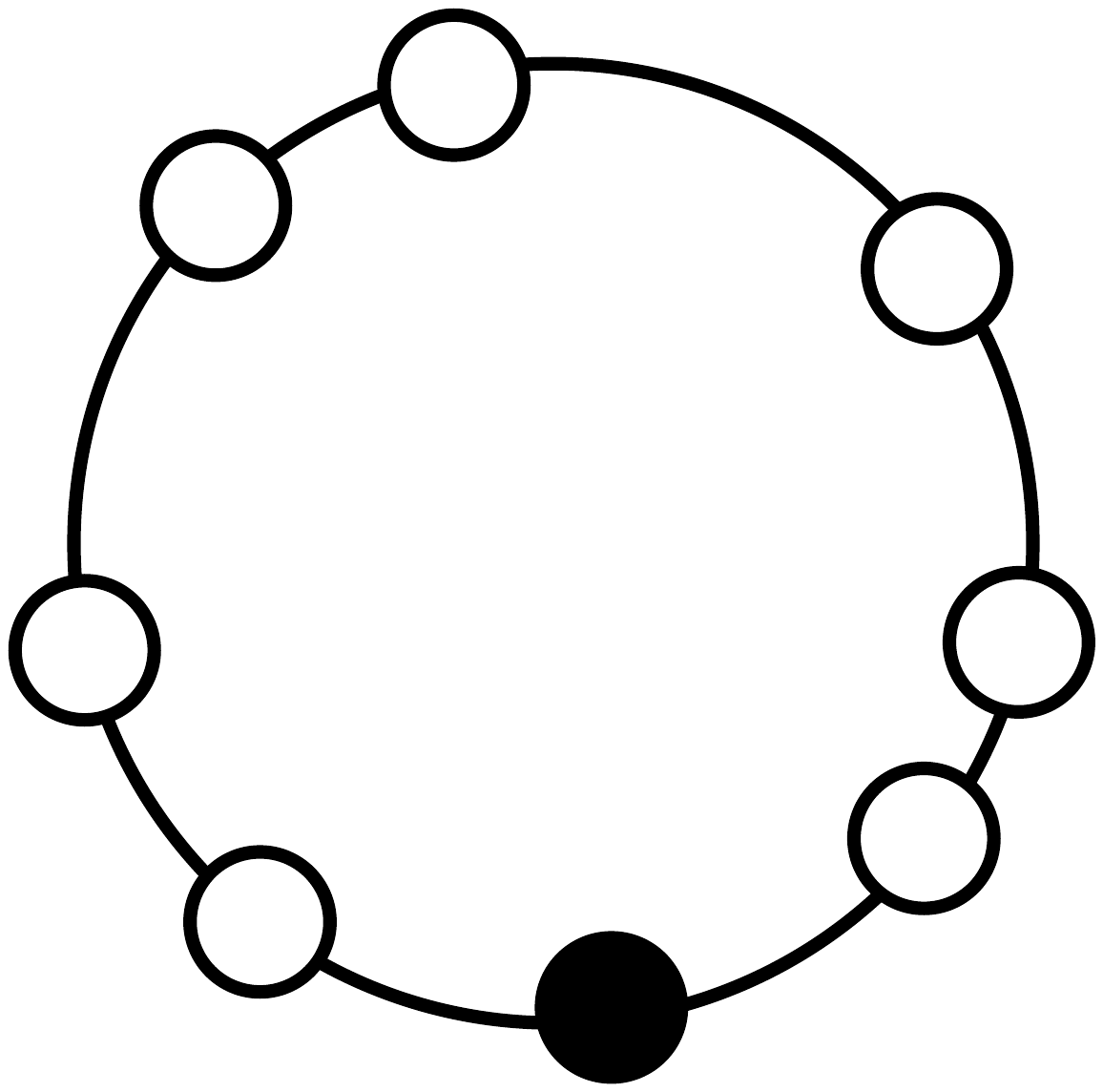}
\caption{\label{model_fig}  A finite one-dimensional model of Bose fluid (BF).
Bose particles are represented by an empty circles.  An additional impurity
particle is represented by the filled circle.  The calculation of the BF and the impurity
excitation spectra is the subject of this paper.}
\end{figure}

Given such toy model, its exact Hamiltonian can be written in the
secondary quantization form.  The BF Hamiltonian adapted to one dimension is
\begin{equation}
\label{H_BF}
\hat{H}_{bf}=\sum_k\frac{k^2}{2m}\hat{a}_k^+\hat{a}_k+
\frac{1}{2L}\sum_{k_1,k_2,l}U_{l}~\hat{a}^+_{k_1+l}\hat{a}^+_{k_2-l}
\hat{a}_{k_1}\hat{a}_{k2}~,
\end{equation}
where $U_l=\int U(x)e^{ilx}dx$ is the Fourier transform of the
pairwise interaction potential, $m$ is the mass of Bose particles,
and $L=2\pi R$ is the length of the ring. The summations over
momenta $k$ are carried out over all possible $k = 0, \pm 1/R, \pm 2/R, ...~$.
Throughout this work we use a simple unit system
in which $\hbar = 1$, all other variables are dimensionless
and are of the order of unity.
The total number of Bose particles, $N$ is an input in our
calculations.  The majority of the calculations are carried out for a liquid like system,
where the average distance between particles is similar to the
range of the pair potential, $N/L \sim 1$. Since the
interaction potential is smooth we do not expect large momenta to play an
important role.  It is verified in all our calculations that results
are well converged when the momentum space is truncated by $-N/R<k<N/R$.

We will consider this model both with and without impurity.  An
impurity Hamiltonian is that of a free particle interacting with
the Bose particles and can be written in coordinate representation for the impurity,
\begin{equation}
\label{H_imp_x}
\hat{H}_{imp}=-\frac{1}{2M}\frac{\partial^2}{\partial
x^2}+\frac{1}{L} \sum_{k,l}V_l~e^{ilx}~\hat{a}^+_{k-l}\hat{a}_k~.
\end{equation}
For the quantum calculations it is more convenient to express this Hamiltonian in momentum representation,
\begin{equation}
\label{H_imp_p}
\hat{H}_{imp}=\sum_p\frac{p^2}{2M}|p\rangle\langle
p|+\frac{1}{L}\sum_{p,k,l}V_l~|p+l\rangle\langle p| ~ \hat{a}^+_{k-l} \hat{a}_k~,
\end{equation}
were $|k\rangle=(1/L)^{1/2} e^{ikx}$ is the momentum wavefunction of impurity,
$M$ is the mass of the impurity and $V_l$ is the
Fourier transform of the impurity/Bose particle pair potential.

Another form of the system Hamiltonian that proves to be useful
is its expression through the Fourier components of the density operator,
\begin{equation}
\label{density}
\hat{\rho_k} = \sum_j e^{ikx_j} = \sum_l \hat{a}_{l+k}^+ \hat{a}_l~.
\end{equation}
The BF Hamiltonian (\ref{H_BF}) can then be written as
\begin{equation}
\label{H_BF_density}
\hat{H}_{bf} = \sum_k\frac{k^2}{2m}\hat{a}_k^+\hat{a}_k+
\frac{1}{2L}\sum_k U_k ~ \hat{\rho}_k ~ \hat{\rho}_{-k}~,
\end{equation}
and the impurity Hamiltonian (\ref{H_imp_x}) becomes
\begin{equation}
\label{H_imp_density}
\hat{H}_{imp}=-\frac{1}{2M}\frac{\partial^2}{\partial
x^2}+ \frac{1}{L} \sum_k V_k ~ e^{ikx} \hat{\rho}_{-k}~.
\end{equation}

\section{Classical Dynamic, Demonstration of Superfluidity}

Even though full quantum calculations are obtained in the following sections,
it is instructional to examine the classical dynamics of this system first.
We notice that the Bose creation/annihilation operators of the system are equivalent to a set of
oscillator degrees of freedom, {\it i.e.}~$\hat{H}=\frac{k^2}{2m}\hat{a}_k^+\hat{a}_k$ is
the hamiltonian of a harmonic oscillator with the frequency
$k^2/2m$.  The dynamics of such degrees of freedom can be considered classically. The interaction term presents a complex unharmonic
coupling between the oscillator degrees of freedom.   The formal use of
Hamiltonian dynamics for $H_{bf}+H_{imp}$ results in the following
equations of motion for the Bose liquid:
\begin{eqnarray}
\label{hequation1}
\dot{x}&=&\frac{p}{M}~,\\
\label{hequation2}
\dot{p}&=&\frac{\partial H(\{a_k^*,a_k\},x,p)}{\partial x}~,\\
\label{hequation3}
\dot{a}_k&=&-i\frac{\partial H(\{a_k^*,a_k\},x,p)}{\partial a_k^*}~,\\
\label{hequation4}
\dot{a}_k^*&=&i\frac{\partial
H(\{a_k^*,a_k\},x,p)}{\partial a_k}~,
\end{eqnarray}
where the function $H(\{a_k^*,a_k\},x,p)$ can be obtained by
replacing operators $\hat{a}_k^+$ and $\hat{a}_k$ by complex variables $a_k^*$
and $a_k$ respectively, {\it i.e.}
\begin{eqnarray}
\label{hamclass}
H(\{a_k^*,a_k\},x,p)=\sum_k\frac{k^2}{2m}a_k^*a_k+\frac{1}{2L}\sum_{k_1,k_2,l}U_{l}a^*_{k_1+l}a^*_{k_2-l}a_{k_1}a_{k2}
+\frac{p^2}{2M}+\sum_{k,l}V_le^{ilx}a^*_{k-l}a_k~.~
\end{eqnarray}
The variables $a$ and $a^*$ can be related to the classical momenta and coordinates
via $a_k=\frac{1}{\sqrt{2}}(x_k+ip_k)$ and
$a_k^*=\frac{1}{\sqrt{2}}(x_k-ip_k)$.  The use of complex variables is convenient for
such a Hamiltonian. The equations of motion, Eqs.~(\ref{hequation1}-\ref{hequation4}), are
equivalent to the Hamilton's equations given that Eq.~(\ref{hequation4}) is the complex
conjugate of Eq.~(\ref{hequation3}).
Notice that in spite of the word "classical" this dynamics is
not written in coordinates of particles. With the exception of the
impurity particle the dynamics is carried out in the space of occupation
numbers.   Thus, the total number of particles is a variable.
However, total momentum and total number of
particles can be shown to be rigorously conserved. These equations
of motion are in fact equivalent to the Gross-Pitaevskii
equations \cite{pitaevskii1998}.

An important drawback of this theory is that the energy minimum does not depend on the
interaction strength and is achieved
by placing all of the Bose particles in the $k=0$ state.
An analysis of the Hamiltonian Eq.~(\ref{hamclass}) shows that $H_{bf}({a_k,a_k^*})$
has the global minimum at $a_0=a_0^*=\sqrt{N}$, and
$a_k=a_k^* = 0$ for $k\not=0$. The latter was verified by performing an extensive
numerical search.  As a result such theory deals with rather unphysical state of the BF and the
classical results may deviate significantly from quantum calculations.

Fig.~2 shows the motion of the particle for the different parameters
of the Hamiltonian Eq.~({\ref{hamclass}}) with $R=3.0$.  The initial conditions for these
trajectories are taken to be $x=0,~p=p_0,~a_0=a_0^*=\sqrt{N}$, $N=19 \sim L$.
Of course, without any interactions the solution is a
straight line $x(t)=p_0t/M$, shown by a dashed line in both plots as a reference.  If the
interaction between impurity and Bose gas is turned "on" but the
interaction part of the bose Hamiltonian is zero (the case of ideal Bose gas with
pair potential parameters $\alpha = 0$, $\beta=0.5$)
the scattering of impurity is clearly observed in the solution
as shown in the top plot.  The scattering is rather complicated and
despite the ordered initial conditions results in a diffusion like motion
after $t \sim 5$. The most important result is shown in the bottom plot.
The Hamiltonian now includes all of the interaction terms
with the different impurity/Bose interaction strengths.  The motion
appears to be nearly that of a free particle, except the velocity of
the particle is different from the free particle value $p_0/M$.
The plot shows two curves with different particle/BF interaction strength,
$\beta=0.3$, and $\beta=1.0$.  The stronger interaction leads to a wavy motion
of the particle.  The oscillation of energy between the particle
and BF can be attributed to the choice of initial conditions for BF,
$n_0=N$, $n_k=0$ for $k\not=0$, which are not
in the equilibrium with the moving particle.  This
result clearly illustrates the microscopic superfluidity of such
system.  The particle cannot transfer energy to the Bose system,
yet the interaction with Bose liquid causes the particle to have an
effective mass $M_{eff}=p_0/\langle dx(t)/dt\rangle$.

\begin{figure}
\label{x_t}
\includegraphics[width=14.0cm]{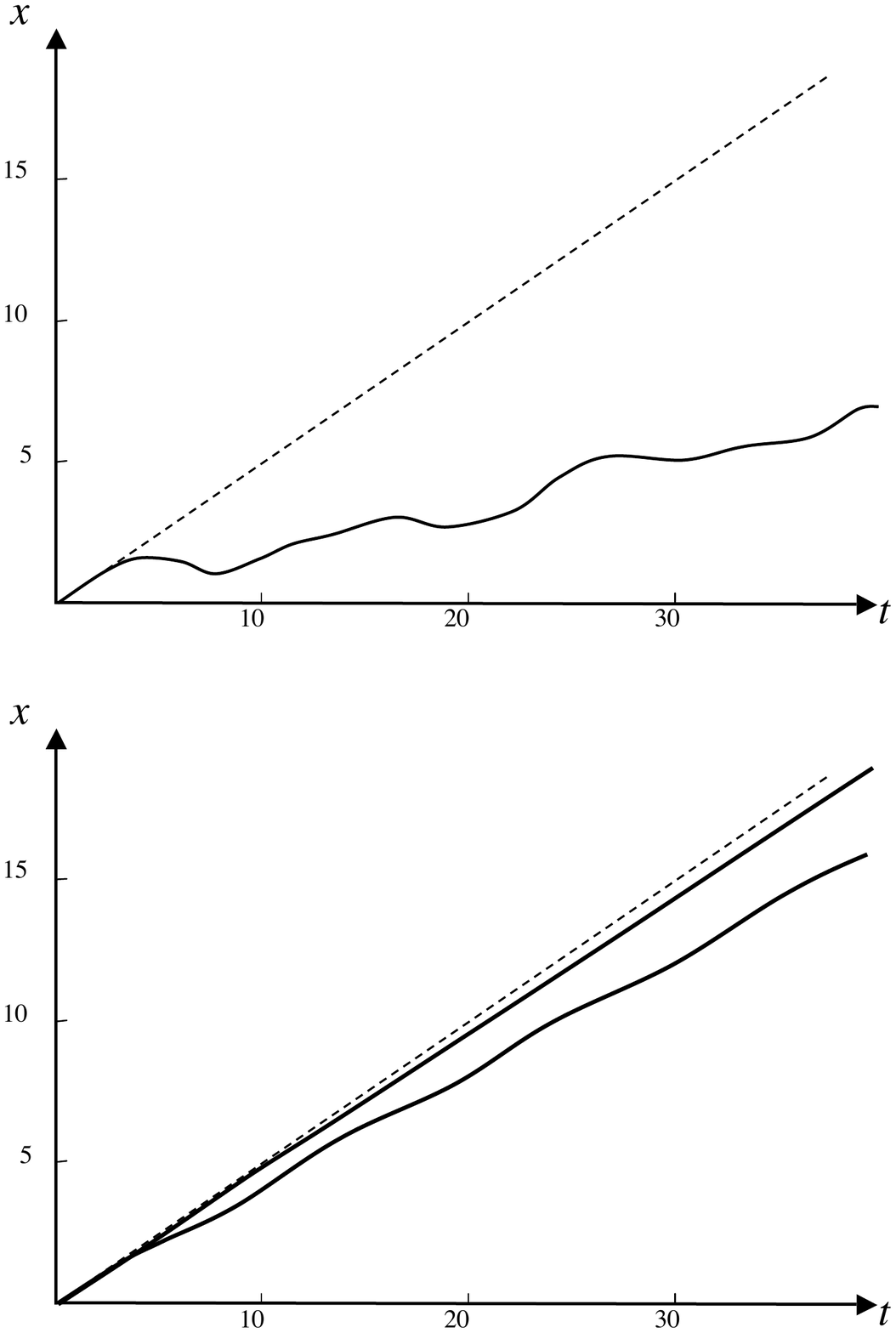}
\caption{ \label{x_t}.  Classical trajectories of impurity interacting with BF obtained by solving
Eqs.~(\ref{hequation1}-\ref{hequation4}).  The dashed line represents the free-particle dynamics.
Top plot shows the motion of impurity interacting with an ideal Bose Gas ($\alpha = 0$).
The bottom plot shows the motion impurity through the interacting BF ($\alpha=0.5$);
two curves are the trajectories that correspond to two different impurity/BF
interaction parameters ($\beta=0.3$ and $\beta=1.0$).}
\end{figure}

To better understand the dynamics observed in these calculations let
us first consider the dynamics of a pure Bose system
(Eqs.~(\ref{hequation3},\ref{hequation4}) with $V_l=0$). Of course, the
minimum of the Hamiltonian function is the stationary solution of
equations of motion. Let us now consider a small deviation of one of
the momentum coordinates from zero $a_k\not=0$ and $a_k^*\not=0$. Doing so
one obtains coupled equations of motion for $a_k$ and $a_k^*$
\begin{eqnarray}
-i\dot{a}_k&=&\left(\frac{k^2}{2M}+\frac{n_0U_k}{L}\right)a_k+\frac{n_o}{L}U_ka_{-k}^*~,\\
i\dot{a}_k^*&=&\left(\frac{k^2}{2M}+\frac{n_0U_k}{L}\right)a_k^*+\frac{n_o}{L}U_ka_{-k}~.
\end{eqnarray}
The simple diagonalization of these equations leads to the two
normal modes that have the frequency given by the Bogoliubov
spectrum
\begin{equation}
\label{bogspectrum}
\epsilon(k)=\sqrt{\left(\frac{k^2}{2M}+\frac{n_0U_k}{L}\right)^2-\left(\frac{n_0U_k}{L}\right)^2}~.
\end{equation}
Thus the "classical" frequency spectrum of the BF normal modes is
the Bogoliubov's spectrum.  Note, that although we used the condensate
number of particles, $n_0 = a^*_0 a_0$, in the above equations it is essentially the
same as the total number of particles, $n_0 \sim N$, due to the chosen initial conditions.

A particle moving with a velocity $v$ exerts the time dependent
driving force on every of the normal modes of the BF given by
\begin{equation}
\label{force}
F_k\sim\frac{n_0}{L}e^{ikx(t)}=\frac{n_0}{L}e^{ikvt}~.
\end{equation}
This force is off-resonance with every normal mode of the BF if $kv<\epsilon(k)$,
{\it i.e.}~the Landau criterion is satisfied.  Nevertheless,
this force results in the forced oscillations of the
normal modes
giving rise to rescaling of the particle velocity.  The energy at
given velocity is the energy of the particle and the total energy of
all forced oscillations of the Bogoliubov's excitations.

This classical picture is rather simple and should translate
directly into the language of perturbation theory of quantum
mechanics. However, the main
difference between the Gross-Pitaevskii equations of motion and
quantum mechanical solution turns out to be the ground state of the BF.
The minimum of the classical Hamiltonian is characterized by all
particles being in the $k=0$ mode.  In the vicinity of
this point the spectrum can only depend on the interaction with
$k=0$ mode. In quantum mechanics this cannot be the case.  The
discrepancy can be thought to arise from non-commutation of the $\hat{a}_k$ and
$\hat{a}_k^+$ operators or the zero point vibrations of Bogoliubov's
excitations. This makes an anharmonic interaction term rather large even for a
single quantum excitation of the BF.

\section{Excitation Spectrum of Pure Bose Fluid}

In this section the full quantum mechanical problem of pure Bose fluid is solved
by direct diagonalization of Hamiltonian Eq.~(\ref{H_BF}).
Since this Hamiltonian commutes with the operator of total momentum,
it is sufficient to consider the blocks of the full Hamiltonian at every total momentum $k$.
The Hamiltonian matrix is constructed in a given basis set and its blocks at each $k$ are
diagonalized.  The analysis reveals which one of
the eigenstates (often, but not always, the lowest energy state) corresponds to the
elementary excitation. Its energy is chosen to be the excitation energy
$E(k)$. The calculations are then repeated for all relevant values of $k$.

Two schemes were implemented to perform such calculations.  The
first approach directly employs the space of occupation numbers as a
basis set.  Basis states are given as $|\{n_k\}\rangle$  We then truncate
the highly excited states and retain only the states that have $n_0>
n_{min}$, where $n_{min}$ is a chosen parameter.  The convergence of
the results with $n_{min}$ is verified by changing this number.
The implementation of this scheme leads to rather expensive
calculations and the convergence was achieved only for weakly
interacting systems ($\alpha \sim 0.1$) or very small systems
(on the order of 10 momentum modes total).

The second approach involves a use of a significantly more efficient
basis.  Knowing that the normal modes of the classical dynamics
are the Bogoliubov's excitations we develop a basis set
based on the Bogoliubov's excitations rather than the occupation
numbers of momentum wavefunctions.  Let us use the main property of
the Bose condensate, the fact that $\langle n_0\rangle$ is the
macroscopic number (or simply large in the case of a finite system).

The first step is to introduce new set of creation and annihilation
operators
\begin{eqnarray}
\label{A1}
\hat{A}_k&=&\frac{1}{\sqrt{n_0}}\hat{a}_0^+\hat{a}_k~,\\
\label{A2}
\hat{A}_k^+&=&\frac{1}{\sqrt{n_0}}\hat{a}_0\hat{a}_k^+~.
\end{eqnarray}
These operators do not actually change the number of particles but
rather promote particles from $0$ to $k$ momentum states.  The
commutation rules for these operators can be easily obtained to be
\begin{equation}
[\hat{A}_k,\hat{A}_{k^\prime}^+]=\left(1+n_k/n_0\right)\delta_{k,k^\prime}\sim\delta_{k,k^\prime}~.
\end{equation}
which reduces to the regular Bosonic commutation in the limit of
large $n_0$. We can then insert the identity operator $\hat{I}=\hat{a}_0^ + \hat{a}_0/n_0$
to rewrite the Hamiltonian via the new operators as
\begin{eqnarray}
\label{hamnew}
\hat{H}&=&\frac{U_0N^2}{2L}+\sum_{k\not=0}\left[\left(\frac{k^2}{2M}+\frac{n_0U_k}{L}\right)
\hat{A}_k^+\hat{A}_k+\frac{n_0U_k}{2L}\left(\hat{A}_k^+\hat{A}_{-k}^+ + \hat{A}_k
\hat{A}_{-k}\right)\right]
\\
&+&\frac{\sqrt{n_0}}{L}\sum_{k\not=l\not=0}U_l\left(\hat{A}_l^+\hat{A}_{k-l}^+
\hat{A}_k+ \hat{A}_k^+ \hat{A}_{k-l} \hat{A}_l\right)
+\frac{1}{2L}\sum_{k_1,k_2,k_3,k_4\not=0}U_{k_1-k_3}
\hat{A}_{k_1}^+ \hat{A}_{k_2}^+\hat{A}_{k_3} \hat{A}_{k_4}\delta_{k_1+k_2-k_3-k_4}~.\nonumber
\end{eqnarray}
Note that we effectively excluded the creation/annihilation of the
particles with $k=0$.  The sums in the Hamiltonian go over all
momenta $k\not=0$ and the Hamiltonian breaks into terms that are
quadratic, cubic, and fourth-oder in these operators.   The
quadratic terms, that constitute the Bogoliubov's Hamiltonian,
consist of the kinetic energy and the part of the
interaction that is proportional to $n_0$.  The cubic terms are
proportional to $\sqrt{n_0}$ and the quartic terms are of the
order of unity.

The next step is to diagonalize the quadratic part of the
Hamiltonian.  Following Bogoliubov we introduce the new operators using
relationships,
\begin{eqnarray}
\label{a1}
\hat{A}_k=\frac{1}{\sqrt{1-L_k^2}}(\hat{B}_k+L_k \hat{B}_{-k}^+)~,\\
\label{a2}
\hat{A}_k^+ = \frac{1}{\sqrt{1-L_k^2}}(\hat{B}_k^+ +L_k \hat{B}_{-k})~.
\end{eqnarray}
The coefficient $L_k$ is chosen such that quadratic part
becomes diagonal.  As it is well known this procedure leads to the
Hamiltonian,
\begin{equation}
\label{H_compute}
H=\sum_{k}\epsilon(k)\hat{B}_k^+ \hat{B}_k +
H^{(3)}\big(\{ \hat{B}_k^+, \hat{B}_k\}\big) + H^{(4)}\big(\{\hat{B}_k^+,\hat{B}_k\}\big)~,
\end{equation}
with the spectrum $\epsilon(k)$ described by Eq.~(\ref{bogspectrum}) and the
coefficients
\begin{equation}
L_k=\frac{L}{n_0U_k}\left(\epsilon(k)-\frac{k^2}{2M}-\frac{n_0U_k}{L}\right)~.
\end{equation}
We do not neglect any of the higher order terms which, as
calculations show, have significant effect on the spectrum.
The terms that are cubic and 4-th order in operators $B_k,B_k^+$ are
represented by $H^{(3)}$ and $H^{(4)}$, respectively. They can be
obtained by substituting Eqs.~(\ref{a1},\ref{a2}) into
Eq.~(\ref{hamnew}). For the sake of space we do not explicitly write
down those obvious but lengthly expressions.

It readily verified that the ground state of the quadratic part of Hamiltonian
Eq.~(\ref{H_compute}) is given by
\begin{equation}
|0\rangle_B=\exp\left[\sum_{k}L_k\left(\hat{A}_k^+ \hat{A}_{-k}^+
+ \hat{A}_k \hat{A}_{-k}\right)\right] |0\rangle~,
\end{equation}
where $|0\rangle$ is the fully condensed ground state of ideal Bose gas ($n_0 = N$).
This is the state with no Bogoliubov's excitations.
The basis set is then generated by the action of Bogoliubov's
creation operators onto this state.
\begin{equation}
| n_{k_1}, ...~ ,n_{k_l} \rangle = \frac{1}{\sqrt{n_{k_l}!}} \left(\hat{B}_{k_l}^+\right)^{n_{k_l}} ...
\frac{1}{\sqrt{n_{k_1}!}} \left(\hat{B}_{k_1}^+\right)^{n_{k_1}} |0\rangle_B~.
\end{equation}
The numerical code generates a list of such basis states and sorts it according to
their total momentum.  A truncation of this basis at the maximum of 5 excitations is
shown to produce converged results for a wide range of system parameters.
A code then uses Hamiltonian Eq.~(\ref{H_compute}) to build the
Hamiltonian matrices for each total momentum $k$ of the system.
These matrices are then
diagonalized resulting in a set of eigenenergies, $E_0^{(k)}, E_1^{(k)}, ...~$, and corresponding
eigenstates, $|\psi_0^{(k)}\rangle, | \psi_1^{(k)}\rangle, ...~$.

At low total momentum $k$ the lowest energy,  $E_0^{(k)}$, is the energy of
the elementary excitation; however, at higher $k$ this is not true, as the elementary
excitation energy should approach $k^2/2m$ and become higher in energy than multiple
excitations that combine to the total momentum $k$.  Thus one needs a criterion to select
an elementary excitation energy out of all eigenvalues.  A simple strategy is to use the
value of the overlap $ | \langle \psi_i^{(k)} | \hat{B}_k^+ | 0 \rangle_B |^2$;  however, for
the strongly interacting  BF all of such overlaps become very small as both ground and
excited states deviate strongly from the Bogoliubov's states.   A physics suggests another
criterion that is used in this work.  By examining the interaction terms in the Hamiltonians, Eq.
(\ref{H_BF_density},\ref{H_imp_density}), one realizes that excitations should be classified
according to the overlap,
\begin{equation}
\label{overlap}
| \langle \psi_i^{(k)} | \hat{\rho}_k | \psi_0^{(0)} \rangle |^2.
\end{equation}
Indeed, this matrix element governs participation of an excited state in most physical
processes.  Note that the right hand side of this expression is precisely the state used by Bijl
and Feynman to develop the theory of helium excitation spectrum  \cite{feenberg}.

The calculated spectrum is shown in Fig.~3.  The parameters of the system are taken be:
$R=3.0$, $N=19$, $\alpha=0.5$, and $m=1$.
The Bogoliubov's spectrum Eq.~(\ref{bogspectrum}) as well as the $k^2/2m$ curve are shown
for comparison. The deviation of the spectrum from the Bogoliubov's result is
significant at low $k$. At small
momenta ($k \le 5/3$) the lowest energy state coincides with the
elementary excitation.  At intermediate momenta the single and double
excitations become strongly mixed.  This results in several states having significant
overlap Eq.~(\ref{overlap}).  Thus several points are shown in Fig.~3 all of which
are the excited states
of the system with momentum  $k$ but they cannot be classified as single or double
excitations.  Finally, at large momenta  the elementary
excitations approach the $k^2/{2m}$ limit of a single particle moving
with the momentum $k$ independently of BF.

\begin{figure}
\label{BF_spec}
\includegraphics[width=17.0cm]{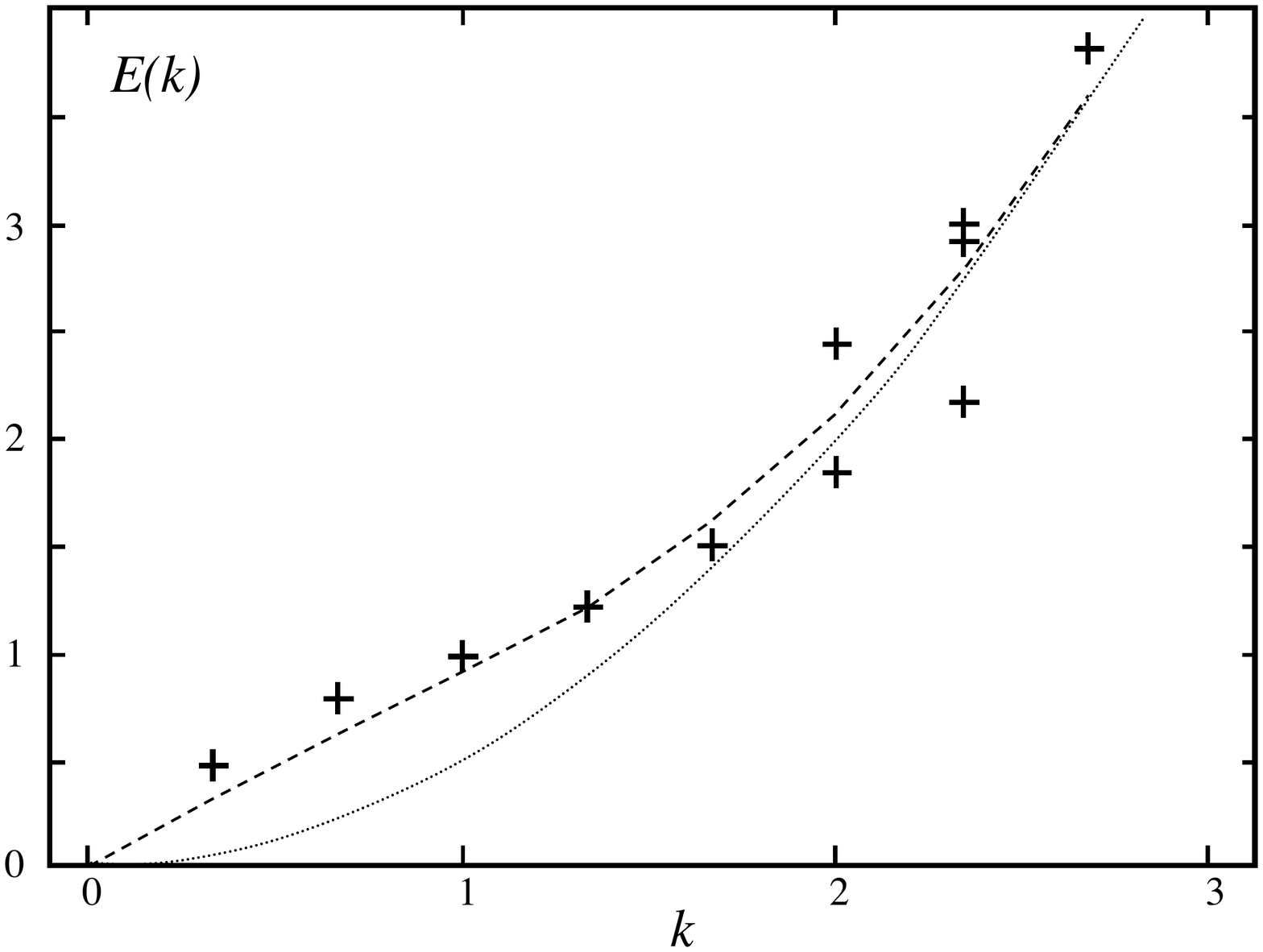}
\caption{ \label{BF_spec}  Crosses show the calculated excitation spectrum, $E(k)$, of
pure BF obtained by full quantum mechanical solution of Hamiltonian
Eq.~(\ref{H_BF}).  The Bogoliubov's excitation spectrum, Eq.~(\ref{bogspectrum}),
(dashed line) and $k^2/2m$ curve (dotted line) are shown for comparison. }
\end{figure}

Despite the fact that the system is one-dimensional its excitation spectrum clearly shows
linear rise at small momentum, thus this system
should retain superfluid-like properties for the purpose of
impurity dynamics at zero temperature.  The calculated slope of the spectrum at small $k$ is significantly different than that of a Bogoliubov's spectrum, which indicates limited
applicability of the classical mean-field treatment presented in previous Section.  In the next Section we add an
impurity to the BF and compute its energy spectrum using the results of the calculations
presented above.

\section{Energy Spectrum of Impurity}

It is natural to use the eigenstates of the BF obtained in the
previous section in order to calculate the energy spectrum of an
impurity.  Using the momentum representation for an impurity
particle we construct a basis set from the momentum states of an
impurity and eigenstates of the BF found in the previous
section. A basis function for a motion with total momentum $p$
can be written as
\begin{equation}
\label{Basis}
|\Phi^{(p)}_{k,j} \rangle = e^{i(p-k)x} | \psi^{(k)}_j \rangle~.
\end{equation}
Momentum, $p$, in this expression is a good quantum number;
it is distributed between the impurity
moving with momentum $p-k$ and the excited state of BF with momentum $k$.
We truncate the number of eigenstates of
the BF that are used in the actual calculations.
It is verified that the use of the 20 lowest states for each momentum $k$, ($j=1,2, ...~,20$)
is sufficient;  no improvement has been found for larger basis sets.

In order to calculate the matrix of the full Hamiltonian the following matrices of the density
operator are pre-computed and saved as the result of the pure BF calculations described
in the last section,
\begin{equation}
\rho^{(k,l)}_{i,j} = \langle \psi^{(k)}_i | \hat{\rho}_{k-l} | \psi^{(l)}_j \rangle~.
\end{equation}
Using the Hamiltonian of the impurity in the form of
Eq.~(\ref{H_imp_density}) the Hamiltonian matrices are computed as
\begin{equation}
H^{(p)}_{k_1,i;~k_2,j} =
\langle \Phi^{(p)}_{k_1,i} | \hat{H}_{bf} + \hat{H}_{imp} | \Phi^{(p)}_{k_2,j} \rangle =
\left( \frac{(p-k_1)^2}{2M} + E^{(k_1)}_i \right) \delta_{k_1k_2} \delta_{ij} +
\frac{1}{L} V_{k_1-k_2} \rho^{(k_1,k_2)}_{i,j}~.
\end{equation}
The energy spectrum of impurity $E(p)$ is found by diagonalization of these matrices.
The results are shown in Fig.~4.  The parameters of the system are:
$R=3.0$, $N=19$, $\alpha=\beta=0.5$, and
$M=m=1$. The $k^2/{2M}$ spectrum of the free particle is shown for comparison.
The dotted line is the
$k^2/{2M_{eff}}$ curve that is fitted to the first calculated point, $p=1/R$.  This leads to
$M_{eff} = 1.29$, compared to $M=1$.
This line passes through the second calculated point with the 0.2\%
accuracy, after which the deviation of the calculated spectrum from the effective mass approximation becomes visible.

\begin{figure}
\label{Imp_spec}
\includegraphics[width=17.0cm]{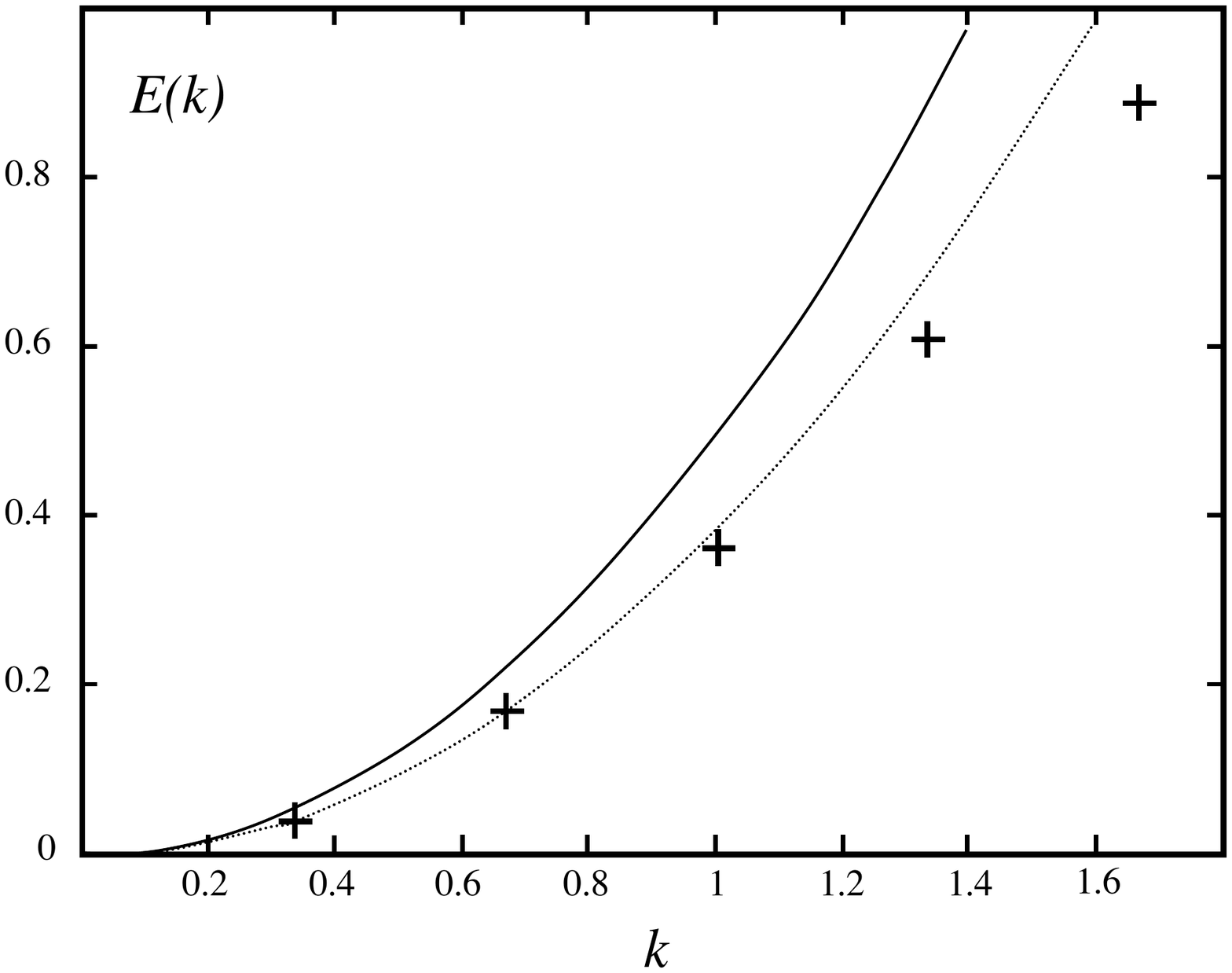}
\caption{ \label{Imp_spec} Crosses show the calculated excitation spectrum of an impurity
particle, $E(k)$. The solid line is the $k^2/2M$ curve, with $M=1$. The dotted line is the
$k^2/2M_{eff}$ fit to the spectrum. $M_{eff} = 1.29$ is obtained using the
lowest point on the spectrum. }
\end{figure}

\section{Relation of Excitation Spectra to the Structure of the Ground State}

Computational results presented in previous sections clearly demonstrate
the microscopic superfluidity in the finite model system.  The excitation spectrum of an impurity
consists of true eigenstates that are well separated from the rest of the
excited states of the system at small $k$.
Interestingly, the finite nature of the model does not have any effect on the physics
except for the quantization of momentum.
The initial several points of the impurity excitation spectrum are in nearly perfect
agreement with the quadratic fit, {\it i.e.}~the effective mass approximation.
Our calculations show that this behavior is true regardless
of the parameters of the system such as density of Bose particles and the strength
of their interaction.  These results appear to be a successful numerical experiment.  While this
experiment clearly shows microscopic superfluidity in a finite system, the results do not
directly give insight  into the underlying physics beyond the Bogoliubov's approximation.
It is clear, however, that the latter
gives the poor estimate of the effective mass as well as poor physical
description of the BF.  In this section we present a microscopic theory
that explains the nature of the BF spectrum as well as an impurity energy based on the
structural properties of the ground state.

A successful theory of the superfluid helium excitation spectrum has been build on the
assumption that the properties of the ground state govern the energy of the excitations.
In particular, Feynman argued \cite{feynman1954}
that given the unknown ground state wavefunction $| \psi_0 \rangle$,
the most natural wavefunction of the excited state with momentum $k$ is given by
\begin{equation}
\label{Feynman_wavefunction}
|\psi_k \rangle=\hat{\rho_k} |\psi_0 \rangle~.
\end{equation}
He then used the properties of the ground state to evaluate the average energy of the
BF in such state.  The first order approximation to the energy spectrum becomes
\begin{equation}
\label{Feynman_spectrum}
E(k) = \frac{k^2}{2m} \frac{1}{S(k)}~,
\end{equation}
where S(k) is the ground state structure factor defined by
\begin{equation}
\label{BF_structure}
S(k) = \frac{1}{N} \langle \psi_0 | \hat{\rho}_{-k} \hat{\rho}_k | \psi_0 \rangle~.
\end{equation}
The structure factor can be related directly to the radial distribution function (RDF), $g(r)$,
of the BF as
\begin{equation}
\label{g_r_BF}
g(r) = 1 - \sum_k e^{ikr} \left( S(k) - 1 \right).
\end{equation}

This theory can easily be verified using the results of the calculation of section IV.
The RDF, Eq.~(\ref{g_r_BF}), calculated using the computed ground state of BF
is shown in Fig.~5.
It is interesting to note that the for the particular choice of system parameters the repulsion
of the Bose particles is clearly observed in the RDF as a well at $r=0$;  however,
the probability of two particles occupying the same volume is significant, in fact the
density of BF at zero distance from a particle is ($\rho(0)/\langle \rho \rangle = 0.37$).
As we found out, the calculations become prohibitively expensive when the
displacement of BF by an atom becomes complete, {\it i.e.}~$\rho(0) \rightarrow 0$, because
the states with large number of Bogoliubov's excitations contribute
significantly to the ground state.
The Bijl-Feynman spectrum is shown in Fig.~6 and compared to the calculated one.
As in the case of the superfluid helium, the spectrum is in excellent agreement at low
values of $k$.
At intermediate momenta the Bijl-Feynman expression
overestimates the excitation energy.
The reason of this effect is clear and can be understood from numerical
results:  the wavefunction Eq.~(\ref{Feynman_wavefunction}) becomes strongly mixed with
double excitations thus lowering the real excitation energy.

\begin{figure}
\label{RDF}
\includegraphics[width=14.0cm]{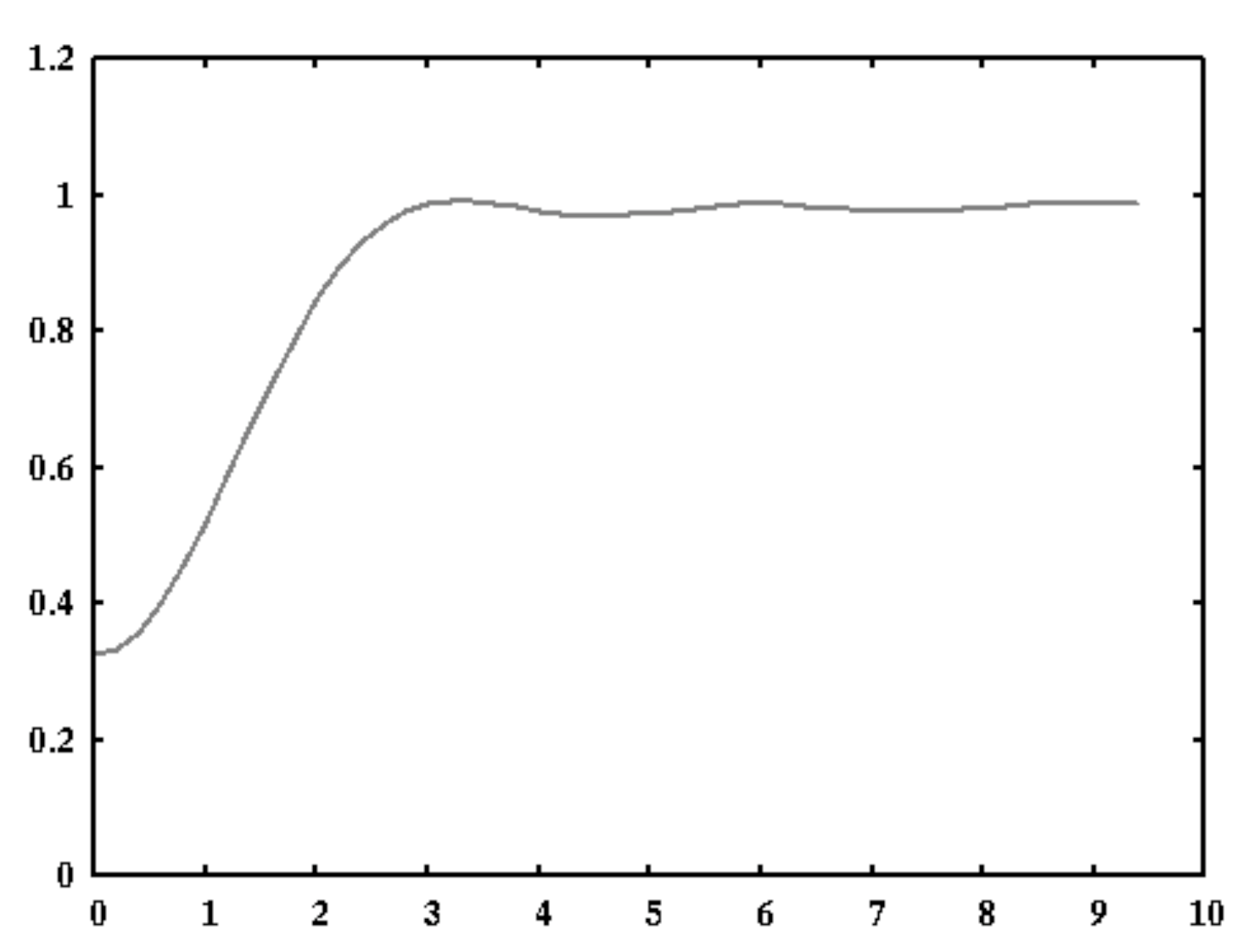}
\caption {\label{RDF}  Radial distribution function, $g(r)$ of the BF computed using the
ground state calculated in section IV. }
\end{figure}

\begin{figure}
\label{BF_excit_struct}
\includegraphics[width=17.0cm]{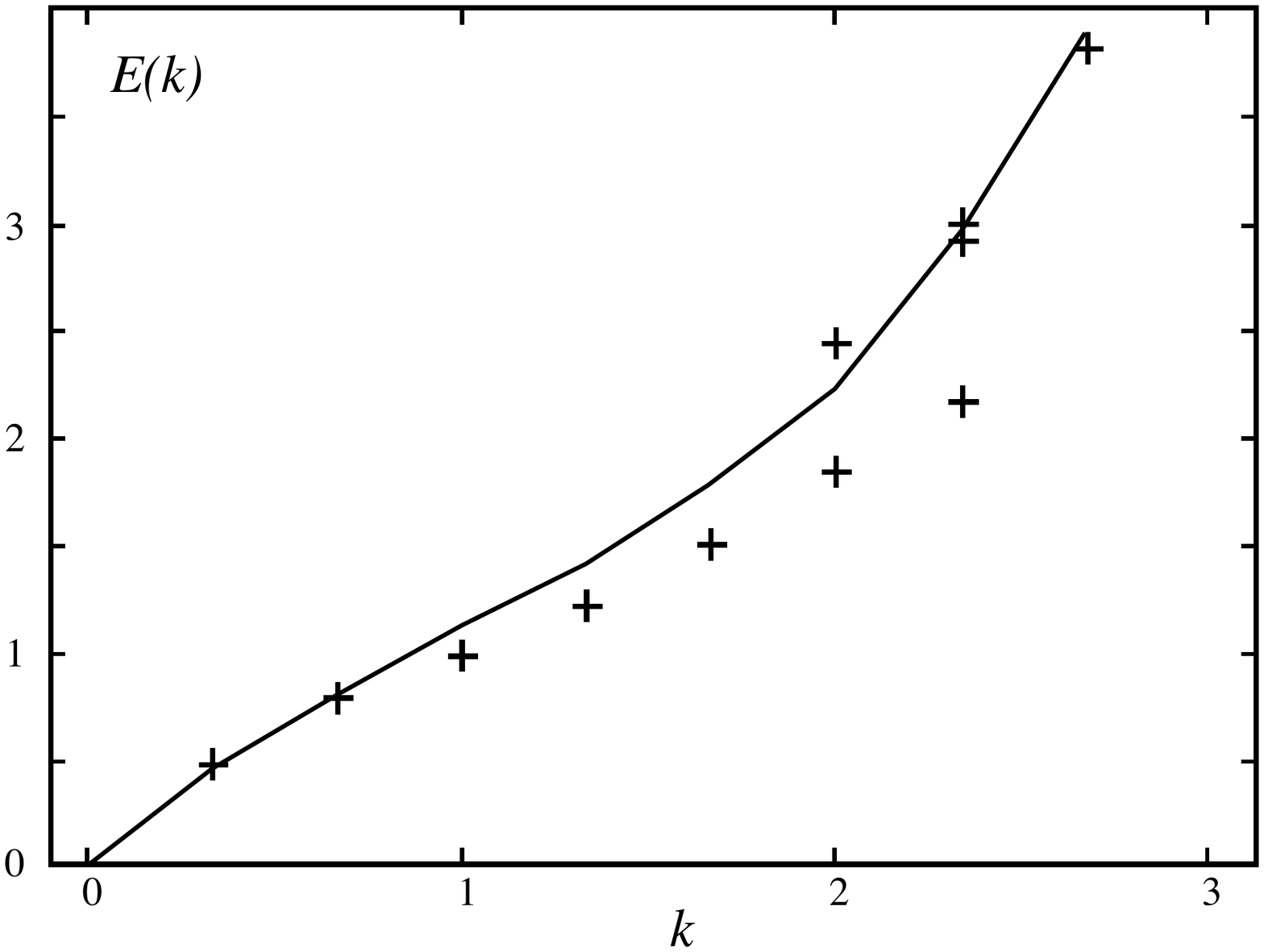}
\caption {\label{BF_excit_struct}  The Bijl-Feynman spectrum, $E(k) = k^2/[2mS(k)]$, (solid line) is
compared to the spectrum calculated in section IV.  $S(k)$ is the structure factor of the
ground state computed in section IV.}
\end{figure}

Similar theory has been developed for the excitation spectrum of impurity \cite{fabrocini1986}.
Let us assume that $|\Phi_0 \rangle$ is the ground state of the combined impurity-BF system.
Wavefunction that represents motion of impurity independent of BF
is given by
\begin{equation}
| \Phi^{(0)}_k \rangle = e^{ikx} | \Phi_0 \rangle~.
\end{equation}
As expected, the variational energy of this wavefunction gives no correction
to the mass of the particle resulting in the energy,
\begin{equation}
\label{E_0}
E^{(0)}_k = \frac{k^2}{2M}~.
\end{equation}
The next step is to incorporate the possible excitations
of the BF into the motion of impurity.  While the exact wavefunctions of BF excited states
are not known it is easy to incorporate the Feynman-like excitations by using the states
\begin{equation}
\label{Phi_l}
| \Phi^{(l)}_k \rangle = \frac{1}{\sqrt{NS(l)}} e^{i(k-l)x} | \hat{\rho}_l \Phi_0 \rangle~.
\end{equation}
The pre-factor $(NS(l))^{-1/2}$ can be shown to normalize this
wavefunction.  The state of Eq.~(\ref{Phi_l}) has the total momentum $k$, which results from
a particle moving with momentum $k-l$ and BF state of momentum $l$.  These
states provide a very natural basis set for the moving particle giving the expression for
the moving particle as
\begin{equation}
\label{expand}
| \Phi_k \rangle = \sum_l c_l | \Phi^{(l)}_k \rangle
\end{equation}
Such states are not orthogonal and their overlap can be evaluated; however, it is does
not enter the perturbation expression derived below.  An important property of these states
is that the relevant Hamiltonian matrix
elements can be easily evaluated and expressed through
the structural properties of the ground state.  Using the method developed by Feynman
one obtains
\begin{equation}
\label{diag_ME}
\langle \Phi^{(l)}_k | \hat{H} | \Phi^{(l)}_k \rangle =
\frac{(k-l)^2}{2M}+\frac{l^2}{2mS(l)}~,
\end{equation}
and
\begin{equation}
\label{off_diag_ME}
\langle \Phi^{(l)}_k | \hat{H} | \Phi^{(0)}_k \rangle =
\frac{kl}{2M} \frac{G(l)}{\sqrt{NS(l)}}~.
\end{equation}
Here we introduced the impurity/BF structure factor
\begin{equation}
G(k) = \langle \Phi_0 | e^{-ikx} \hat{\rho}_k | \Phi_0 \rangle~.
\end{equation}
Using Eqs.~(\ref{diag_ME},\ref{off_diag_ME}) the second order perturbation
correction to the energy Eq.~(\ref{E_0}) is obtained
\begin{equation}
\label{perturb}
E^{(2)}_k = \sum_l
\frac{ \left(\frac{kl}{2M}\right)^2 \frac{G(l)^2}{NS(l)} }
{\frac{k^2}{2M} - \frac{(k-l)^2}{2M} - \frac{l^2}{2mS(l)}}~.
\end{equation}
At small $k$, the main term of the denominator Eq.~(\ref{perturb}) is the
energy of the BF excitation $l^2/[2mS(l)]$, thus this equation can be reduced to
\begin{equation}
E^{(2)}_k \approx - \frac{k^2m}{M^2} \frac{1}{N}  \sum_I \left( \frac{G(l)}{S(l)} \right)^2~,
\end{equation}
which leads to the effective mass of impurity given by
\begin{equation}
\label{M_eff}
M_{eff} = M \left[1-\frac{m}{M} \frac{1}{N}  \sum_l \left( \frac{G(l)}{S(l)} \right)^2 \right]^{-1}~.
\end{equation}

Once again, this theory can be easily tested using our numerical results.  The impurity/BF
structure factor $G(k)$ is essentially a Fourier transform of the BF density distribution around
the particle.  Both functions are obtained by the analysis of the ground state obtained in our
calculations.  The density distribution of Bose particles around an impurity, given by
\begin{equation}
\rho(r) = \frac{1}{L} \left( N  - \sum_k G(k)e^{ikr} \right)~,
\end{equation}
is shown in Fig.~7.  This function is similar to the RDF of BF, except is not normalized to unity at
large distances but instead represents the actual density of the BF.  In our particular case the
density of Bose particles is increased in the region away from impurity compared
to the unit value because of  the actual displacement of particles in the small system.
The energy spectrum predicted by Eqs.~(\ref{perturb}) is compared to the
calculated spectrum in Fig.~8.  The result shows nearly perfect numerical agreement with the first
two points of the spectrum and starts to deviate at larger $k$.  The main source of the deviation
can be identified as the interaction with double excitations not considered by Eq.~(\ref{expand}).
The value of effective mass predicted by Eq.~(\ref{M_eff}), $M_{eff}/M = 1.27$, is
very similar to the value $1.29$
computed directly.  It is interesting to note, that the effective mass cannot be understood
simplistically as the mass of BF displaced by the interaction with impurity
particle.  As our calculations show, $M_{displaced} \sim 2$, while $M_{eff} - M = 0.29$.

\begin{figure}
\label{rho}
\includegraphics[width=14.0cm]{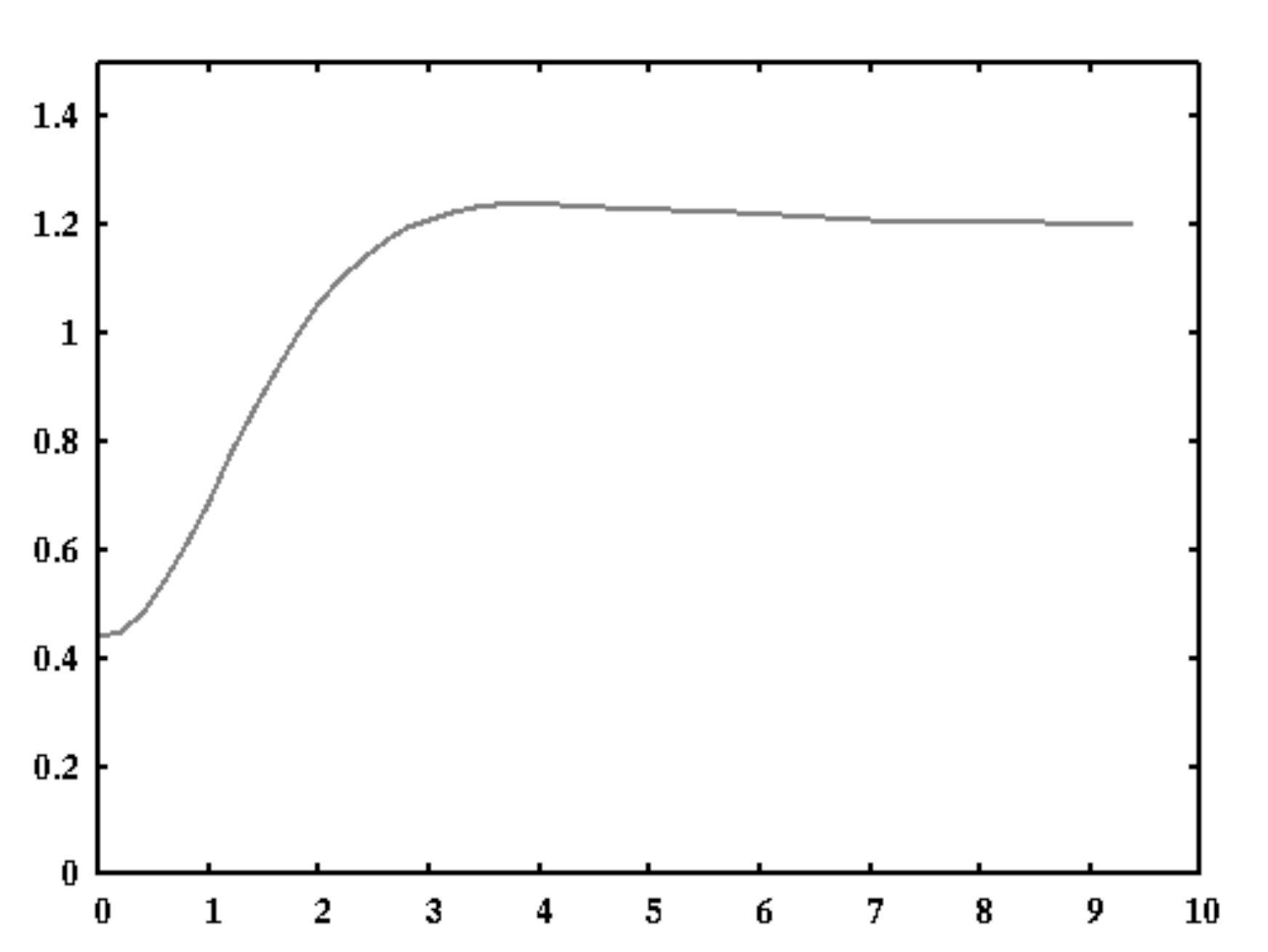}
\caption {\label{rho}  Average density of BF as a function of distance from impurity $\rho(r)$
resulting from the ground state of impurity/BF system computed in section V.}
\end{figure}

\begin{figure}
\label{excit_struct}
\includegraphics[width=17.0cm]{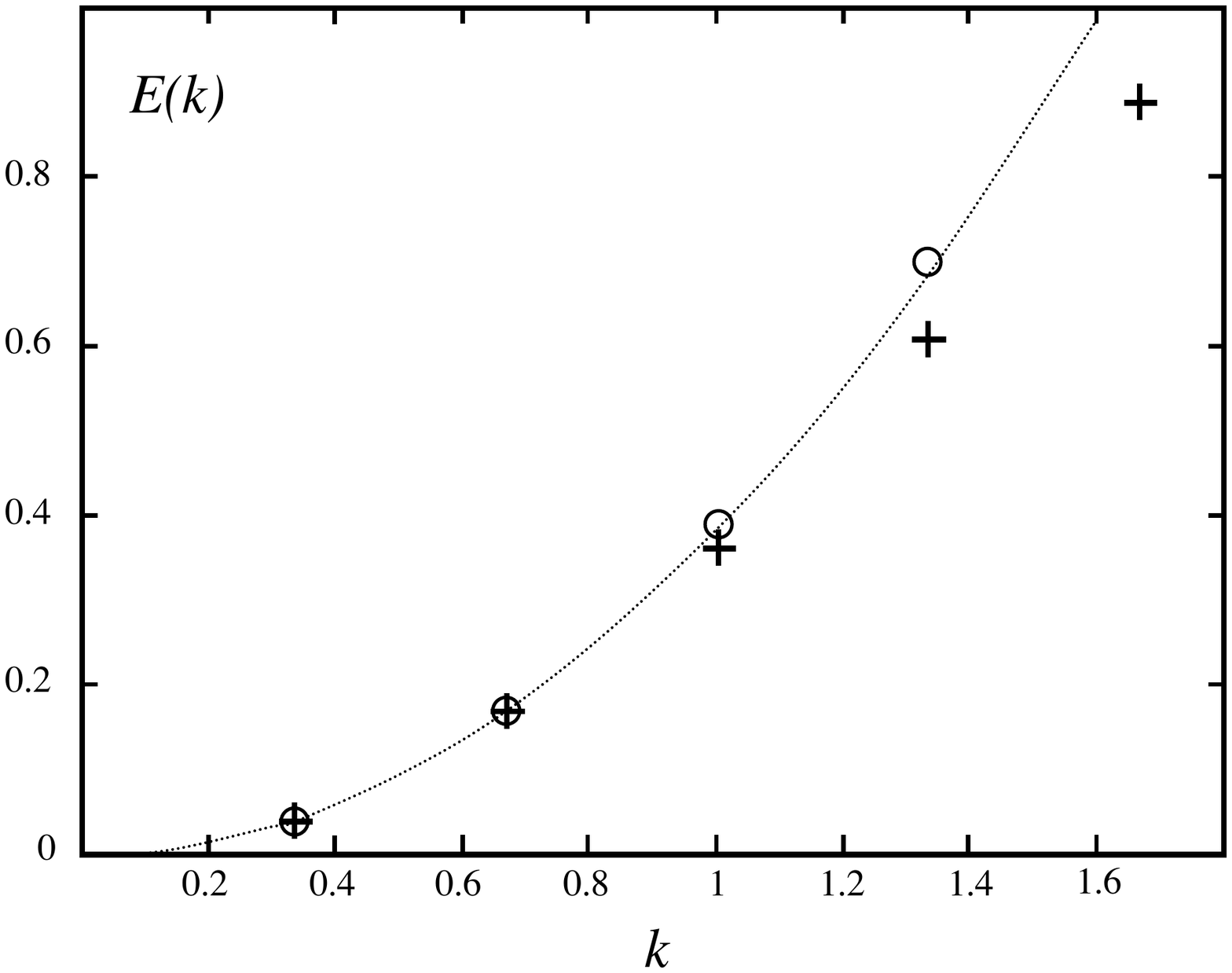}
\caption {\label{excit_struct}  Impurity excitation spectrum calculated using Eq.~(\ref{perturb}) (circles) compared to the spectrum calculated in section V (crosses).  Dotted line is the effective
mass approximation. }
\end{figure}

The perturbation theory expression, Eq.~(\ref{perturb}) is equivalent to the one-phonon intermediate perturbation (OIP) result obtained in ref. \cite{fabrocini1986}.
Although this equation is not as accurate for a more realistic system of $^3$He in $^4$He, our calculations show that it captures the essential physics
of impurity motion.  For our system this result gives an accurate description of impurity
excitation spectrum at small $k$.  The most important feature of this
theory is that excitation energies are expressed entirely through the structure of the
ground state.  The latter can be computed for a realistic system such as liquid He using the
wealth of the imaginary time path integral techniques. In other words, the dynamical problem
is reduced to a statistical one.  A rather complete effort in this direction has been
recently undertaken by Zillich and Whaley \cite{whaley2004z}.  Their work combines the
variational CBF theory with the diffusion Monte Carlo simulations to compute moments of
inertia of molecules in helium droplets.  Our work shows quantitative success of much
simpler approximations based entirely on the density distributions.
This success raises hope that
simpler methods based on the perturbation expansions could be used to solve similar
problems.

\section{CONCLUSIONS}
\label{conclusion}

In this work we present a computational study of a finite one-dimensional system
that consists of a BF and an additional impurity particle confined to a ring.
None of the interactions are taken to to be small yet the full quantum calculations are
converged for this system.  The effects of the strong particle repulsion is clearly observed
in the structure of the ground state.  It is shown that this system exhibits a property of microscopic
superfluidity, {\it i.e.}~there is a brunch of energy states that corresponds to
the motion of an impurity particle.
These are true energy eigenstates of the system that are well separated from the
continuum of quantum states of the BF.  The calculations suggest a most natural
analytical description of impurity motion.  In particular, the excited states of the impurity
can be obtained from the ground state of the impurity-BF system.  It is shown that the
excitation spectrum can be accurately predicted from the structure of the ground state.

We believe that the calculations presented in this
work provide a valuable illustration to the analytical theory of superfluidity developed
over the last sixty years.  The classical dynamics nearly ideally illustrates the
work of Bogoliubov exemplifying the main approximation inbuilt in his treatment.
The dynamics of this system clearly shows the validity of the Landau criterion and
shows that it is applicable to finite systems.  Finally, the full quantum calculations
illustrate in detail the microscopic theory of superfluidity developed by Feynman
and his successors.  It is our hope that a complete understanding
of this model system will help us to combine the modern state of the art
computational methods with the wealth of analytical theory developed decades ago.

\section{acknowledgments}
This work has been supported by the NSF CAREER award ID 0645340.

\newpage

\end{document}